\documentclass[sigplan,screen]{acmart}
\AtBeginDocument{%
  \providecommand\BibTeX{{%
    \normalfont B\kern-0.5em{\scshape i\kern-0.25em b}\kern-0.8em\TeX}}}

\copyrightyear{2023} 
\acmYear{2023} 
\setcopyright{acmlicensed}\acmConference[HILDA '23]{Workshop on Human-In-the-Loop Data Analytics}{June 18, 2023}{Seattle, WA, USA}
\acmBooktitle{Workshop on Human-In-the-Loop Data Analytics (HILDA '23), June 18, 2023, Seattle, WA, USA}
\acmPrice{15.00}
\acmDOI{10.1145/3597465.3605228}
\acmISBN{979-8-4007-0216-7/23/06}

\usepackage{xcolor}%
\usepackage{tabularx}

\begin{document}

\title{Data Makes Better Data Scientists}

\author{Jinjin Zhao}
\email{j2zhao@uchicago.edu}
\affiliation{%
  \institution{University of Chicago}
  \city{Chicago}
  \state{IL}
  \country{USA}
}

\author{Avigdor Gal}
\email{avigal@technion.ac.il}
\affiliation{%
  \institution{Technion -- Israel Institute of Technology}
  \city{Haifa}
  \country{Isreal}
}

\author{Sanjay Krishnan}
\email{skr@uchicago.edu}
\affiliation{%
  \institution{University of Chicago}
  \city{Chicago}
  \state{IL}
  \country{USA}
}


\begin{abstract}
With the goal of identifying common practices in data science projects, this paper proposes a framework for logging and understanding incremental code executions in Jupyter notebooks. This framework aims to allow reasoning about how insights are generated in data science and extract key observations into best data science practices in the wild. In this paper, we show an early prototype of this framework and ran an experiment to log a machine learning project for 25 undergraduate students.

\end{abstract}

\maketitle

\section{Introduction}
A good data scientist must be able to reason about uncertain data, how this data interacts with code, and how to incorporate extensive domain knowledge with this data to draw her insights~\cite{dhar2013data}. The confluence of skills needed for even simple data science tasks makes it hard to run disciplined evaluations in data science~\cite{battle2022behavior}. For example, consider a new software tool that claims to improve data scientist productivity. Should productivity be measured in terms of a reduction of lines of code or the time spent working? Furthermore, how should these metrics be reconciled with questions about accuracy and robustness? 

While many other industries have been revolutionized by ``Metric-based Management'' ~\cite{bauer2004kpis}, ironically, the key performance indicators (KPIs) of data science productivity have not yet been established at the individual or group level. Admittedly, there are numerous studies on human decision-making with data in psychology~\cite{edwards1954theory}, data visualization~\cite{battle2018evaluating}, data science education~\cite{kreuter2019data}, and business decision-making~\cite{h2014strategists}. 
However, there is limited information on the processes and tasks that take place in real-world data science workflows that might cut across multiple toolkits and involve multiple data scientists.  Notable seminal projects in this space include SQLShare dataset~\cite{jain2016sqlshare} and the KGTorrent dataset~\cite{quaranta2021kgtorrent}.

We argue that these studies are limited since they miss crucial steps in the data exploration and code design process -- and only log the final state of a query or a code cell respectively. Code that fails to execute or is deleted can provide crucial clues about the data, data model, and/or data scientist. For example, a confusing, ill-documented schema may result in a large number of erroneous queries. With the goal of improving the community's quantitative understanding of data science processes, we propose a system that tracks incremental code executions in Jupyter notebooks. Jupyter notebooks are computational notebooks, which are collections of cells that intersperse code, results, and narrative. 
The framework tracks all modifications and executions to data science code written in a notebook.


These logs are passed through an analysis module that: (1) recognizes the current task focus of the data scientist based on code patterns, (2) identifies what data assets are being manipulated based on data provenance, and (3) builds a temporal model of the data scientist's overall process.

In this paper, we show an initial experiment that logs the workload of 25 undergraduate students during a machine learning task. As our logging framework matures, we envision several core applications, as follows.

\noindent  \textbf{More Informative User Studies.} We are interested in quantitatively capturing the process by which data scientists generate insights from data. We want to validate common assumptions (e.g., data cleaning is the most challenging part of data science) and identify impactful problems to improve this process.

\noindent  \textbf{More Complex User Studies.} Data science is implemented across practically all domains by diverse groups of data scientists with varying expertise. We want our system to enable workload capture across these different groups and observe differences of approach across domains.

\noindent  \textbf{Data Governance Insights.} For some scenarios, restricting data access to particular use cases and users is extremely important. In these cases, by tracking the workflow process, a record of what data is used for insights and how that data is accessed can be generated. This information can shape policies for data sharing and governance.

\noindent \textbf{Data Science Education. } By understanding the differences in workflows between data science novices and experts, we can quantify the knowledge gained from experience and education. This knowledge can be incorporated to improve our current data science curriculum. We can also generate specific feedback based on a student's workflow on assignments, allowing for individualized instruction without increasing the instructor's burden.

\section{Related Work}\label{sec:related}
\subsection{Jupyter Notebooks} Jupyter notebooks are known to be messy, with poor preservation of history and out-of-order workflows \cite{chattopadhyay_messy, kery_messy}. Accordingly, tools have been developed to generate cleaner code slice views on notebooks for human use \cite{janus, nbsafety, nbslicer, nbgather, lineapy}. All these tools provide the groundwork for fine-grained quantitative analysis of notebook code but do not directly analyze computational behavior.

\subsection{Data Science Studies} Numerous interview studies have been made to profile data science workflows \cite{kery_story, shankar_2022, kandel_2012, whither}. These studies have limitations due to human bias when self-reporting their work \cite{bias}. Quantitative studies on data science and notebook development have primarily focused on the final pipeline \cite{lee_2020, xin_paper_study, adam_2018, pimentel_2019, psallidas_2019, xin_google}. They give overviews on the distribution of code and supporting text and are valuable studies in best practices for reproducibility and final code design. However, they do not provide insight into the process of generating those workflows. 


\section{Architecture and Student Logs} \label{sec:arch}

\begin{figure}[t]
    \centering
    \includegraphics[width=0.8\columnwidth]{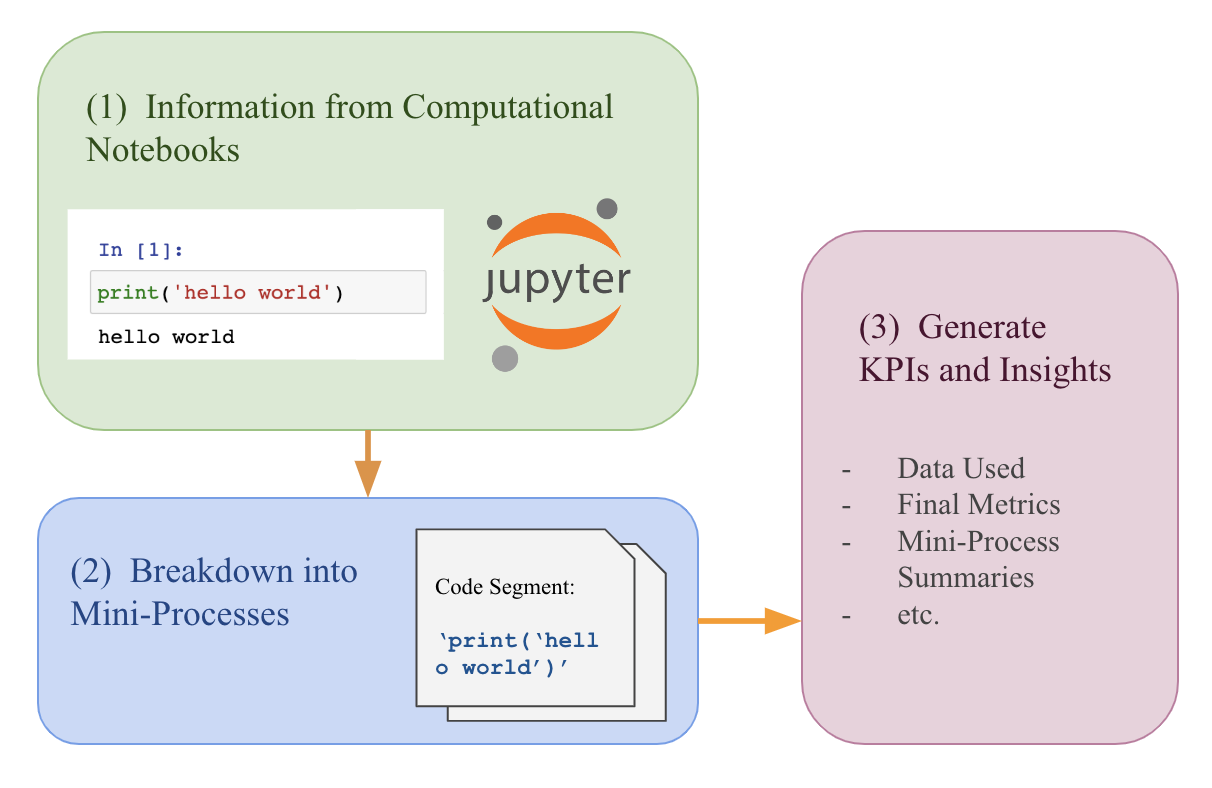}
    \caption{Architecture of proposed Jupyter notebook log framework}
    \label{fig:architecture}
\end{figure}

\subsection{Architecture}
In Figure \ref{fig:architecture}, we outline the envisioned framework for data science logging in Jupyter notebooks. This framework has three steps: (1) logging all code edits to create a timeline of a data science project, (2) a breakdown of this log into self-contained mini-processes that describes the capture of data insights, (3) the extraction of KPIs from the mini-processes that allow for comparisons across processes and projects.

\subsection{Experiment Setup} \label{sec:setup}
An experiment was run to capture data science workflow in a classical machine learning problem. The concrete goal of this experiment is to record detailed logs of a typical amateur data science workflow using Jupyter notebooks.  We recruited 25 undergraduate students from an introductory data science course at the University of Chicago and gave them a standard machine learning task. Given the full 2018 Air Flight dataset~\cite{kaggle_dataset}, they had the task of predicting flight departure delays of over 15 minutes on the 2019 version of the same dataset (with any column containing information at or after departure withheld). The 2018 and 2019 datasets had 5,602,937 and 8,091,684 samples, respectively. 


\begin{figure}[t]
    \centering
    \includegraphics[width=0.8\columnwidth]{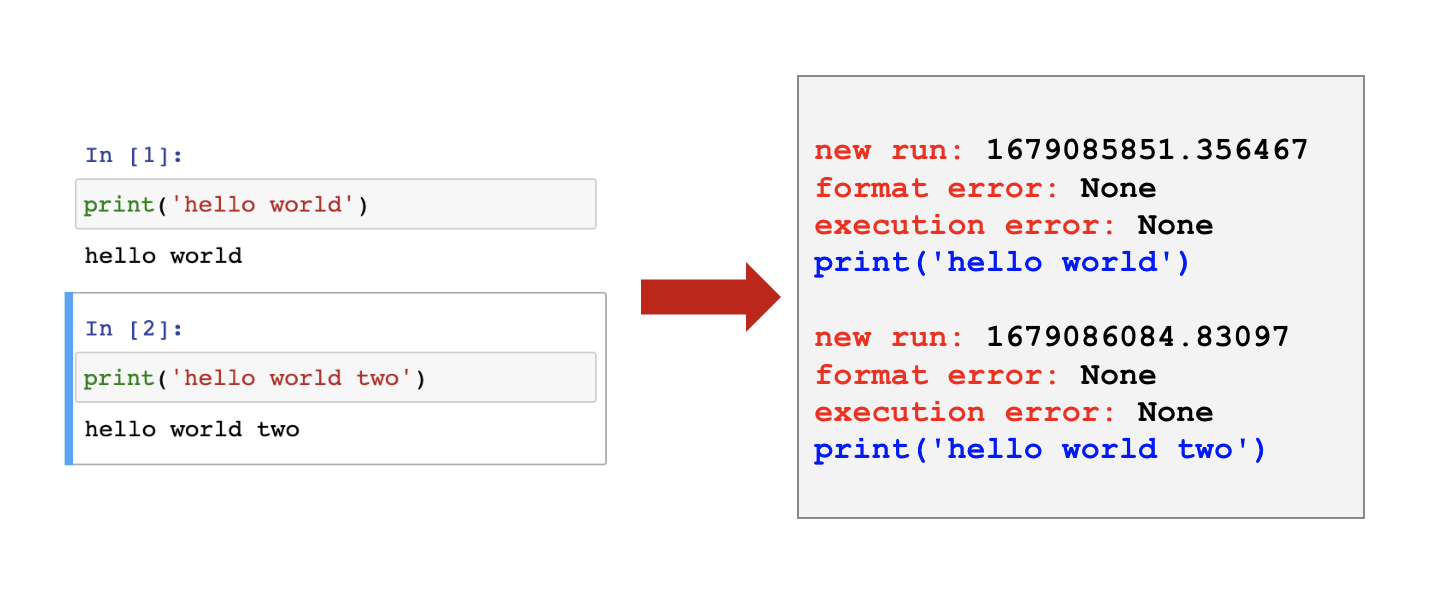}
    \caption{Example of execution log generated from a computational notebook}
    \label{fig:log}
\end{figure}

In order to track this process, we create an execution log with a record for each cell run in a Jupyter notebook. This log is similar to the log in \texttt{nbgather}~\cite{nbgather} and the IPython's default history database. However, there are subtle critical differences in the information tracked. We consider one execution log for each notebook, logging data over the entire Jupyter notebook history. We record cell content, execution time, and error statements for each cell run. The output of the cells is not logged, a design decision due to anecdotal observation of the volume of outputs generated compared to the code written. Figure~\ref{fig:log} shows an example of the execution log for a simple Jupyter notebook.

In addition to the execution log, we have the final notebook submission and a prediction file for flight departure delays of over 15 minutes on the 2019 dataset from each student. To summarize, the sources of information we captured in this experiment are: (1) code for each cell executed throughout the lifetime of the notebook, (2) wall-clock time of execution, (3) any execution results/exceptions generated, (4) final machine learning predictions, and (5) the final submitted Jupyter notebook.


\bibliographystyle{abbrv}
\bibliography{references}

\end{document}